# VIVoNet: Visually-Represented, Intent-Based, Voice-Assisted Networking

Amar Chaudhari, Amrita Asthana, Atharva Kaluskar, Dewang Gedia, Lakshay Karani, Levi Perigo, Rahil Gandotra and Sapna Gangwar

Interdisciplinary Telecom Program, University of Colorado Boulder, USA

### ABSTRACT

*Networks have become considerably large, complex and dynamic. The configuration, operation, monitoring, and troubleshooting of networks is a cumbersome and time-consuming task for the network administrators as they must deal with the physical layer, underlying protocols, addressing systems, control rules, and many other low-level details. This research paper proposes an Intent-based networking system (IBNS) coupled with voice-assistance that can abstract the underlying network infrastructure and allow administrators to alter its behavior by expressing intents via voice commands. The system also displays the real-time network topology along with the highlighted intents on an interactive web application that can be used for network diagnostics. Compared to traditional networks, the concepts of software-defined networking (SDN) make it easier to integrate a voice assistant that allows configuring the network based on intents.*

## 1. INTRODUCTION

Software-defined networking (SDN), as the new networking paradigm, allows for programmability of the different networking layers [1]. The Open Networking Foundation (ONF) defines the three layers of SDN as - the infrastructure layer, responsible for the forwarding of packets, the centralized control layer, responsible for adding the flow entries at the infrastructure layer, and the application layer, consisting of the program that communicates requirements to the control layer [5]. The OpenFlow protocol works at the southbound interface (SBI) between the controller and forwarding elements, while API's such as REST are employed at the northbound interface (NBI) allowing applications to communicate with the controller.

Intent can be defined as a high-level policy that determines the behaviour of a network. Intent is 'what' the user wants from the network, without worrying about 'how' it is done. The Intent-based networking system (IBNS) allows the administrator to enforce these high-level policies which are independent from the lower-level details - specific network technologies and vendor specific features. [2]. The time required to change such an intent in a traditional network is substantial [3]. For example, to increase or decrease network bandwidth at a specific time, the administrator would need to modify the configuration of each device in the network at a specific time. Contrarily, an IBNS would automatically modify the network configuration to satisfy the high-level intent, making the network infrastructure agile. So, comparing with the programming paradigm, the traditional method of configuration is imperative while the intent-based method is declarative [4].

Voice-assisted technologies have been used for many years, from automated voice recognition in telephone systems to speech-to-text Dictaphones. But recent advances in machine learning have allowed the development of sophisticated natural language processing algorithms making voice-assisted technologies ubiquitous – Amazon's Alexa, Apple's Siri, Microsoft's Cortana and Google's Assistant. The speed, efficiency, and convenience offered by these devices is leading toward the trend of less screen-interaction [6].

The research work proposed in this paper integrates these different technologies together to develop a system that would assist administrators in network management, operations, and automation. The remainder of the paper is organized as follows: Section 2 provides a review of the existing body of knowledge and the research novelty of VIVoNet. Sections 3 and 4 describe the research overview, and the results and analysis, respectively. Section 5 provides a conclusion and addresses scope for future enhancements.

## 2. RELATED WORKS

### 2.1. Voice-assisted automation

Spoken Dialogue System's (SDS's) are intelligent agents that help users complete a task more efficiently via spoken interaction [7]. SDS's are being integrated into a variety of devices such as smartphones, smart televisions, smart refrigerators and car navigation systems [8]. An SDS can support numerous applications in the enterprise, education, healthcare, entertainment, and government [7], but little work has been done on adding SDS for network management systems. Amrutha et al. [9] proposed a wireless home automation system controlled by a computer. The proposed system automatically translates spoken words into text commands in MATLAB. The text commands are transmitted to a microcontroller which controls the household appliances via their corresponding relays.

Rajalakshmi et al. [10] proposed a system to connect and control the Internet of Things (IoT) devices using voice assistance. The proposed system utilizes Alexa Voice Service to develop a customized skill that is used to connect IoT devices and publish them. Amazon Web Services (AWS) IoT service is employed as a centralized management platform for various IoT devices, and the AWS Lambda service triggers to process voice commands from the user. The proposal provides a use case of controlling IoT devices with voice and suggests possibilities of similar implementations that can be extended to computer networking and data monitoring.

### 2.2. Network management

Simplified network management has been one of the primary motivations behind innovations in networking, starting from efforts in active networking in mid - 90s, to the For CES working group in the early 2000s, with SDN and OpenFlow emerging in late 2000s [1]. The centralized control plane with a global view of the network allows the administrator to manage the network from a single entity [33]. And the decoupling of the control and forwarding planes makes various network management tasks such as automation, configuration management, and traffic engineering less challenging [34]. The VIVoNet system was developed on the principles of SDN to enable programmatic management at each layer of the network – management, control, and forwarding.

Addressing the need for network configuration languages that could translate administrator policies directly on to the underlying infrastructure, policy descriptive languages such as Procera [11] and Frenetic [35] were proposed. In [11], policy definitions are employed to express the changing network conditions and state via network configuration and management. Since the ONF, or the industry, has not standardized the northbound API, we explore network management using voice-assistance in this research.

## 2.3. Visual representation

Monitoring and visualization of traffic is an important aspect of network management [11]. They can greatly aid in a better understanding of the network and assist in the everyday management tasks, such as designing and testing new network services or identifying the traffic patterns. Isolani et al. pointed out the benefits of visualizing and understanding the network topology in an SDN environment [12]. Watashiba et al. implemented an OpenFlow visualization tool that visualizes the network topology and allows administrators to modify flow entries on the switches [13]. Wassapon et al. created a transparent monitoring tool, Opimon, which is placed as a proxy between the controller and the switches, to avoid modifying the controller code [14]. Guimaraes et al. addressed the benefits of visualization for SDN management in terms of improving administrator productivity and reducing costs [15]. VIVoNet employs visual representation to provide feedback to the user to verify that the speech commands implemented the desired changes in the network topology.

## 2.4. Intent-based networking

Projects exist which provide intent capabilities for SDN network management based on high-level application policies. The Open Daylight project has proposed NeMo, Network Modelling for applications, a northbound API for intent driven networking [16]. NeMo provides the capabilities to configure virtual networks using HTTP. The ONOS project also provides an interface with limited set of intent capabilities called the Intent Framework [17].

Han et al. construed the objectives of an IBNS and the roadblocks to its development [2]. Intent composition, conflict checking, mapping, and installation are addressed, and express high-level requirements specifications are reviewed. Cohen et al. discussed two core components of IBNS – abstraction from the underlying vendor-dependent infrastructure, and enforcement of specified behavior on the network environment [18].

Cisco published a white paper on intent-based networking and discussed its advantages – speed, agility, and reduction in human error [19]. They examine the working of such a system which involves three steps:

  i. Translation: Comprehension of required business intent and its translation into code.
 ii. Activation: Deployment of the intent on the network infrastructure.
iii. Assurance: Logging, monitoring and alerting.

Vivo Net extends the fundamental concept behind intent-based networking by extending the intents to include user utterances as well.

## 2.5. Research novelty

After a study of the available literature, no work was found which addresses network management and automation using voice assistance. This research involves the integration of intent-based networking with voice assistance and visual representation to approach network management in a novel way. We believe that the Vivo Net system developed from this research can be of immense use to network administrators, importantly the visually-impaired audience as well. The primary research question we answered is can a network management system be developed that accepts voice commands from the user and dynamically configures intents on the infrastructure in real-time. We identified the following sub-problems to be answered to effectively develop and evaluate our research.

i. Can user utterances be identified and translated into network intents?
ii. Can the different translated intents be configured on the network infrastructure?
iii. Can the configured intents be visually represented for the user in real-time?

## 3. RESEARCH OVERVIEW

For the experiment, a testbed was created which simulates a typical backbone network of a company with points of presence (POPs) in multiple cities (see Fig. 1). All components of VIVoNet are virtualized on physical machines using the VMware ESXi hypervisor [27]. The backbone network contains virtual switches as device nodes. The virtual switches are open-source Open vSwitch (OvS) (v2.5.1) running OpenFlow 1.3 [20]. The system consists of a centralized SDN controller, Floodlight, to manage all the OvS [21]. The POPs that connect to the virtual devices are FRRouting (v4.0) instances, connected in a full-mesh running eBGP and exchanging assigned prefixes [28].

VIVoNet system was developed using the Python programming language. Django (v1.11.10) framework has been used to develop the front-end and the REST APIs [22].

The Intent Engine and Speech Recognition System are modules written in Python for the system. The web server is a Nginx instance that acts as a reverse proxy by accepting HTTP requests from the users and Alexa and forwards them to the Django application via Web Server Gateway Interface (WSGI) [23]. The database consists of open-source MariaDB (v10.3) software running on the server [24].

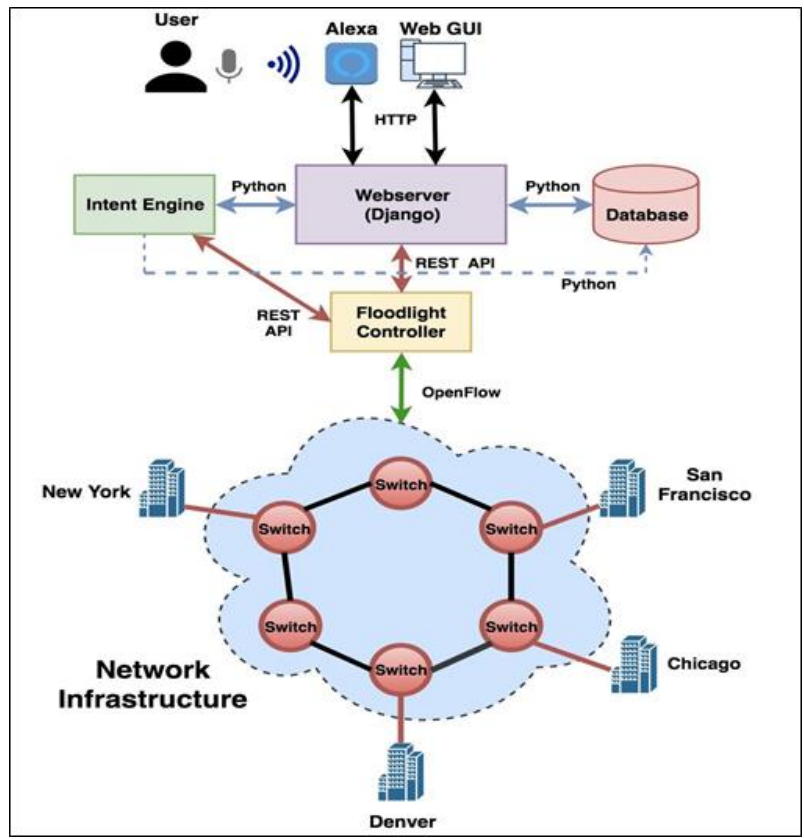

Figure 1. System setup

To feed voice input to the system, Amazon Echo Dot (2nd Generation) was used. The Echo Dot was connected to a Wi-Fi network that has access to Amazon-Alexa service over the Internet. A MacBook Pro running the latest Google Chrome browser was used to access the Web GUI via the Internet. DUO security service was employed to provide two-factor authentication [29].

The following four components were developed to build the VIVoNet system that work in tandem to produce the desired results:

  i. Speech recognition
 ii. Intent processing engine
iii. Infrastructure
 iv. Visual representation

The administrator can direct the network behavior via voice commands. The Speech Recognition system accepts these voice commands via Amazon's smart assistant Alexa. Amazon allows developers to extend Alexa's capabilities by defining custom skills [25]. A custom skill called VIVoNet has been created using the Alexa Skills Kit (ASK). The VIVoNet skill is a collection of sample utterances to invoke specific intents and slot types which help in getting more information from the user.

In the VIVoNet system, the Intent Engine is a Python application responsible for converting user-requested intents into a translated network configuration. It creates appropriate flows, and the controller pushes them out using OpenFlow to the switches. The switches do not have a control plane and forward traffic depending on the flows present in their flow tables. The OpenFlow channels between the controller and switches have been encrypted using SSL. For persistent storage of end-locations, prefixes, live intents and, existing flows, a database is maintained on the front-end of the system. As soon as flows are pushed to the infrastructure, the database is updated with the corresponding values. The Visual Representation application renders the network topology via REST API, retrieves the intents from the database, and highlights the path indicating the specified intents.

There are three intents supported by VIVoNet:

  i. Least latency: Set the path with the minimum delay from source to destination.
 ii. High bandwidth: Set the path with the maximum available bandwidth from source to destination.
iii. Least hop count: Set the shortest path between source and destination based on the number of hops.

The web interface of the VIVoNet system provides a real-time view of the entire network infrastructure. It uses REST API to extract the network topology in a JSON-based format. With the help of VIS.JS, a dynamic browser-based visualization library, the JSON-based network graph is converted into an aesthetic real-time network topology [26].

The web application also provides a fail-over scenario for specifying text-based intents in case the Alexa service is not running (see Fig. 2). The user will specify the intent type, source and destination in the drop-down menu, and the selected values are then converted into a JSON-based format which simply emulates the voice-based command for the Alexa service. The JSON input is sent as a POST request to the Intent Engine which then implements the configuration on the network topology.

Since VIVoNet has the potential to control traffic engineering in the network infrastructure, it is important to protect the system from malicious users. The web application restricts access by allowing only authorized users to log in and request changes. The users accessing the system are first landed on a login page where their credentials are validated from the authentication database. Once verified, the users can then view the network topology information and create intents. Two-factor authentication has also been incorporated into the system. Lastly, the OpenFlow channels have been encrypted with SSL to prevent eavesdropping.

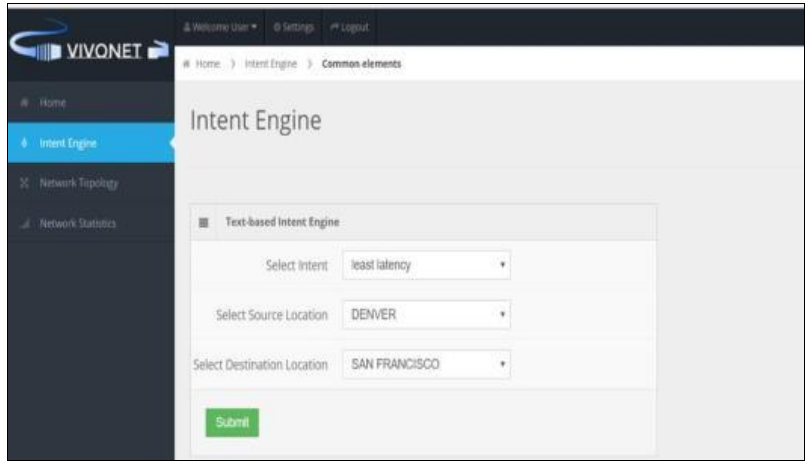

Figure 2. Text-based backup page

## 4. RESULTS AND ANALYSIS

In the experiments, we attempted to answer the three sub-problems stated in section 2 to ultimately answer the primary research question also stated in section 2.

### 4.1. Can user utterances be identified and translated into network intents?

Fig. 3 shows the flow of Alexa's interaction with the user and the VIVoNet system. The user first says a command such as "Alexa, launch VIVoNet." Alexa analyzes the user utterance and identifies the corresponding intent (Create Intent). Alexa then continues the conversation in a funnel format with the aim of filling the configured slot values by asking additional questions.

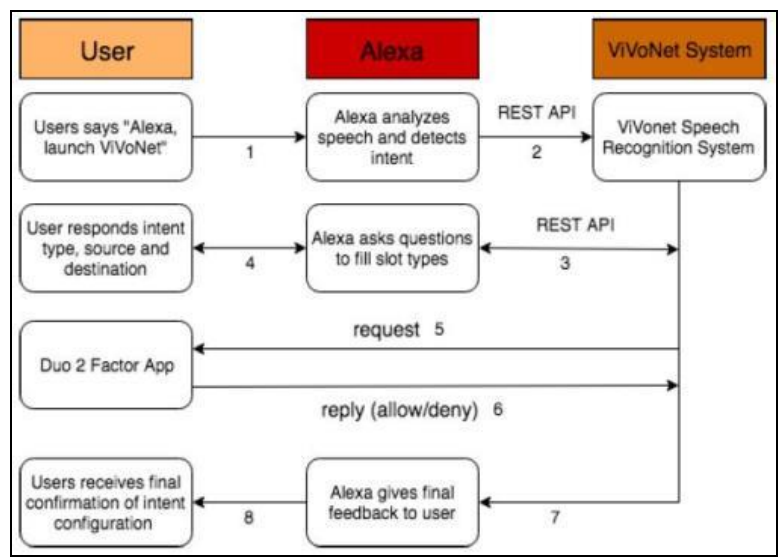

Figure 3. Alexa skill flow

Table 1 shows the utterances that will result in the invocation of a particular intent. For example, if the user says 'launch VIVoNet,' Alexa first converts the speech into text and then matches the text with all the sample utterances from the configured intents. If there is a match, Alexa tries to match user inputs with the slot-types and sends a POST request to the REST API endpoint (VIVoNet/ask/alexa) with these slot-type values and the invoked intent.

Table 1. Sample utterances to invoke intents

| Intents | Sample utterances |
|---|---|
| LaunchRequest | launch VIVoNet |
| CreateIntent | <ul><li>VIVoNet setup a {intent_type} path from {from_city} to {to_city}</li><li>{confirmation}</li><li>{intent_type}</li><li>{from_city} to {to_city}</li></ul> |

Table 2 shows the configured slot-types used by the VIVoNet Skill. The {intent_type} is a custom slot-type with allowed values of least hop-count, least latency, and high bandwidth. Similarly, {from_city} and {to_city} utilize Amazon's in-built slot-types of AMAZON.City which can detect any city name worldwide.

Table 2. Intent schema and slot types

| Slots Name | Slots Type | Sample Slots Values |
|---|---|---|
| {intent_type} | intent_type | least hopcount, least latency, high bandwidth |
| {from_city} | AMAZON.City | Denver, San Francisco (any U.S. city) |
| {to_city} | AMAZON.City | New York, Chicago (any U.S. city) |
| {confirmation} | confirmation | Yes/No |

## 4.2. Can the different translated intents be configured on the network infrastructure?

Fig. 4 shows the process flow of the Intent Engine. The Speech Recognition system provides the required intent and the source and destination cities to the Intent Engine. The Intent Engine retrieves the switch Data Path Identifiers (DPID) from the controller via the REST API and finds all available paths between the two DPIDs by querying the controller again. The controller uses OpenFlow's OFPT_STATS_REQUEST message to fetch the required information from the switches. Taking the intent into consideration, it computes the best path amongst the available paths. The Intent Engine then constructs a flow based on parameters like switch DPID, the input and output port, source prefix and destination prefix. After constructing the flows for all the switches selected in the best path, they are pushed on to the respective switches using the OpenFlow OFPFC_ADD or OFPT_FLOW_MOD messages. Once it is verified the push operation was successful, the intent and flows are written to the database, from where the Visualization System retrieves them and displays on the web interface. The final step in the intent processing is to return a value to the Speech Recognition system – True for success and False for failure.

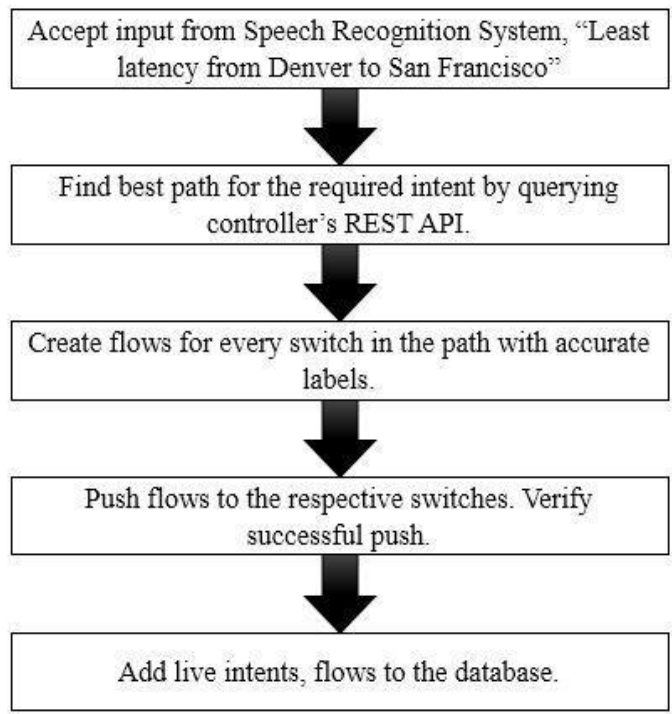

Figure 4. Intent Engine process flow

## 4.3. Can the configured intents be visually represented for the user in real-time?

The Intent Engine stores the configured path for the user specified intent in the database. The web application extracts the configured path using internal APIs and is populated on the web application. The user can select any particular intent from the generated drop-down and view the highlighted intent path in realtime. The visual display of highlighted intents aids the user to see and verify the traffic flow based on the configured intents (see Fig. 5).

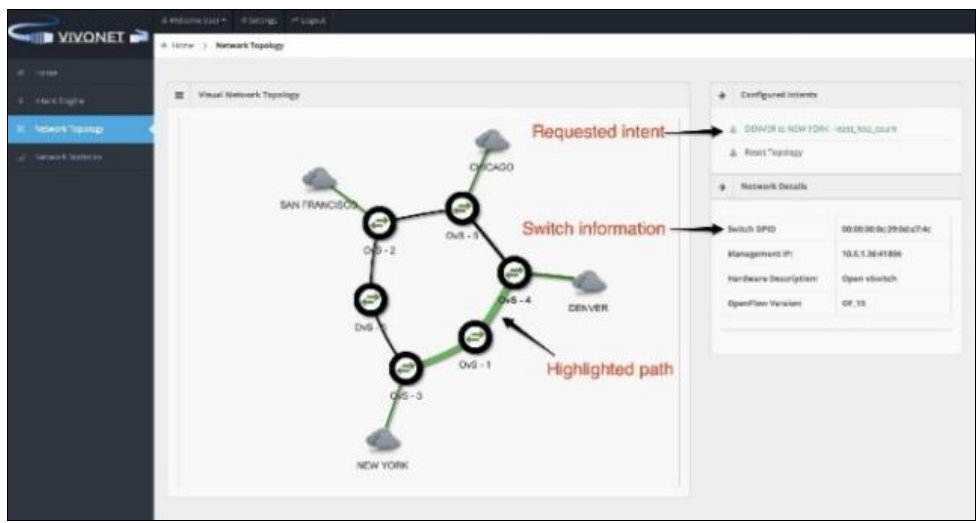

Figure 5. Result displayed on GUI

### 4.4. Analysis

#### 4.4.1. Time saving:

In today's hyper-scale networks, agility is of utmost importance, and optimizing the time for deployments, scaling and debugging is critical. This makes time an invaluable resource for any enterprise. On average, the VIVoNet system required 0.0016 seconds to create, push and verify flows on any one OvS device for one intent. Alexa has a hard timeout value of 8 seconds per request/response. The VIVoNet system has four request/response pairs; therefore, the conversation with Alexa takes 32 seconds at most. Hence, this research estimates a total maximum time of 32.08 seconds to configure a network of five OvS devices.

#### 4.4.2. Capital efficient and flexible:

Most of the components used in the VIVoNet system are available for no cost which reduces the cost of the overall solution and increases its market value. Additionally, the use of open-source solutions enables administrators to customize according to their requirements without being dependent on vendors [32].

#### 4.4.3. Granularity in automation:

SDN has various advantages that can be leveraged to automate management of large-scale networks. SDN has a centralized control plane which can reduce overhead and downtime [30]. Traditionally, MPLS/RSVP-TE along with DSCP values has been used to segregate traffic. However, DSCP provides limited marking capabilities. VIVoNet system utilizes OpenFlow which provides extensive 40-tuple matching fields [31], which enables more granular traffic marking.

#### 4.4.4. Accessibility:

Taking an interdisciplinary approach, a system like VIVoNet can greatly help operators with visual impairment by making information more accessible, thereby allowing them to interact with their infrastructure.

## 5. CONCLUSION

This paper explored an approach for simplifying the configuration of network infrastructures through a northbound interface and the use of voice assistance. In this research, a novel state-of-the-art intent-based networking system using open-source solutions was developed that can apply high-level intents to a network infrastructure via voice commands with real-time visualization. A user-friendly and precise front-end was developed with a rich set of features that can give administrators insight into the network topology, visualization of various live intents, and modular information about the network infrastructure. Security was inculcated by implementing SSL connections in the software-defined infrastructure and including two-factor and login authentication. With the technology moving towards the voice-enabled smart assistants, the VIVoNet system developed from this research has the potential to simplify and automate network management and operations.

## 5.1. Future work

This paper is a prototype for a voice-based intent networking system. The system has a wide range of applications that could simplify network management, and we believe that after adding scalability, maintainability, and monitoring, VIVoNet has the potential to be of great assistance in managing the network of various sized organizations. Additionally, owing to its voice assistance feature, the VIVoNet system has the capability to facilitate visually impaired engineers in network administration.

To enhance the results from this research, further research could be performed to evaluate VIVoNet in the following scenarios:

- Orchestrate networks via Zero Touch Provisioning with voice.
- Retrieve infrastructure health metrics such as CPU utilization, memory, and link flapping using voice commands for troubleshooting.
- Network monitoring by setting alarms.
- Traffic segregation at the application layer.
- Directing automated self-healing networks.

## REFERENCES

[1] N. Feamster, J. Rexford, and E. Zegura, "The Road to SDN: An Intellectual History of Programmable Networks," ACM Queue, New York, NY, USA, Tech. Rep., 2013.

[2] Y. Han, J. Li, D. Hoang, J. Yoo and J. Hong, "An intent-based network virtualization platform for SDN," IEEE 12th International Conference on Network and Service Management (CNSM), pp. 353-358, 2016.

[3] Y. Tsuzaki and Y. Okabe, "Reactive configuration updating for Intent-Based Networking," IEEE International Conference on Information Networking (ICOIN), pp. 97-102, 2017.

[4] Open Networking Foundation, "Intent NBI – Definition and Principles," Technical Recommendation, 2016.

[5] Open Networking Foundation. Software-Defined Networking (SDN) Definition - Open Networking Foundation. [online] Available at: https://www.opennetworking.org/sdn-definition/.

[6] H. Feng, K. Fawaz and K. Shin, "Continuous Authentication for Voice Assistants", ACM 23rd Annual International Conference on Mobile Computing and Networking, pp. 343-355, 2017.

[7] V. Kpuska and G. Bohouta, "Next-generation of virtual personal assistants (Microsoft Cortana, Apple Siri, Amazon Alexa and Google Home)," IEEE 8th Annual Computing and Communication Workshop and Conference (CCWC), pp. 99-103, 2018.

[8] B. Dhingra et al., "Towards End-to-End Reinforcement Learning of Dialogue Agents for Information Access," Association for Computational Linguistics annual meeting, 2017.

[9] S. Amrutha et al., "Voice Controlled Smart Home," International Journal of Emerging Technology and Advanced Engineering, vol. 5, January 2015.

[10] A. Rajalakshmi and H. Shahnasser, "Internet of Things using Node-Red and alexa," 7th International Symposium on Communications and Information Technologies (ISCIT), pp. 1-4, 2017.

International Journal of Computer Networks & Communications (IJCNC) Vol.11, No.2, March 2019International Journal of Computer Networks & Communications (IJCNC) Vol.11, No.2, March 2019

[11] A. Voellmy, H. Kim, and N. Feamster, "Procera: a language for high-level reactive network control," in Proceedings of the 1st Workshop on Hot Topics in Software Defined Networks (HotSDN '12), pp. 43–48, ACM, Helsinki, Finland, August 2012.

[12] P. H. Isolani, J. A. Wickboldt, C. B. Both, J. Rochol and L. Z. Granville, "Interactive monitoring, visualization, and configuration of OpenFlow-based SDN," IFIP/IEEE International Symposium on Integrated Network Management, pp. 207-215, 2015.

[13] Y. Watashiba et al., "OpenFlow Network Visualization Software with Flow Control Interface," IEEE 37th Annual Computer Software and Applications Conference, pp. 475-477, 2013.

[14] W. Wassapon, P. Uthayopas, C. Chantrapornchai and K. Ichikawa, "Real-time monitoring and visualization software for OpenFlow network", 15th International Conference on ICT and Knowledge Engineering (ICT&KE), pp. 1-5, 2017.

[15] V.T. Guimaraes et al., "Improving productivity and reducing cost through the use of visualizations for SDN management," IEEE Symposium on Computers and Communication (ISCC), pp. 531-538, 2016.

[16] NeMo-project.net, NeMo - The Application's interface to Intent Based Networks. [online] Available at: http://NeMo-project.net/.

[17] Wiki.onosproject.org, Intent Framework – ONOS – Wiki. [online] Available at: https://wiki.onosproject.org/display/ONOS/Intent+Framework.

[18] R. Cohen et al., "An intent-based approach for network virtualization," IFIP/IEEE International Symposium on Integrated Network Management, pp. 42-50, 2013.

[19] Cisco. (2018). Intent-Based Networking. [online] Available at: https://www.cisco.com/c/en/us/solutions/intent-based-networking.html.

[20] Openvswitch.org. Open vSwitch | Production Quality, Multilayer Open Virtual Switch. Available at: http://openvswitch.org/.

[21] Project Floodlight. (2018). Projects - Project Floodlight. [online] Available at: http://www.projectfloodlight.org/projects/.

[22] Djangoproject.com. (2018). The Web framework for perfectionists with deadlines — Django. [online] Available at: https://www.djangoproject.com/.

[23] NGINX. (2018). NGINX — High Performance Load Balancer, Web Server, & Reverse Proxy. [online] Available at: https://www.nginx.com/.

[24] MariaDB.org. (2018). MariaDB.org. [online] Available at: https://mariadb.org/.

[25] Developer.amazon.com. (2018). Alexa Skills Kit - Build for Voice with Amazon. [online] Available at: https://developer.amazon.com/alexaskills-kit.

[26] Visjs.org. vis.js - A dynamic, browser based visualization library. [online] Available at: http://visjs.org.

[27] Vmware.com. ESXi | Bare Metal Hypervisor | VMware. [online] Available at: https://www.vmware.com/products/esxi-and-esx.html.

[28] Frrouting.org. FRRouting. [online] Available at: https://frrouting.org/.

[29] Duo.com. Double Up on Security With Two-Factor Authentication (2FA). [online] Available at: https://duo.com/product/trusted-users/two-factor-authentication.

[30] J. Rath, "Integrating SDN into the Data Center." [online]
Available at: http://documents.tips/business/integrating-sdninto-the-data-center.html.

[31] Open Networking Foundation, "OpenFlow Switch Specification; Version 1.3.0(Wire Protocol 0x04).", p.40, 2012. [online].
Available at: https://www.opennetworking.org/images/stories/downloads/sdnresources/onf-specifications/openflow/openflow-spec-v1.3.0.pdf.

[32] R. Gandotra and L. Perigo, "SDNMA: A Software-defined, Dynamic Network Manipulation Application to Enhance BGP Functionality," in Proceedings of the 20th IEEE International Conference on High Performance Computing and Communications (HPCC), 2018.

[33] D. Gedia, and L. Perigo, "NetO-App: A Network Orchestration Application for Centralized Network Management in Small Business Networks" in 4th International Conference on Computer Science, Engineering and Information Technology (CSITY-2018), Sydney, Australia, pp. 61-72, July, 2018, DOI: 10.5121/csit.2018.81106.

[34] D. Gedia, and L. Perigo, "A Centralized Network Management Application for Academia and Small Business Networks" in Information Technology in Industry (ITII)-indexed in web of science, Australia, Sept, 2018.

[35] N. Foster, R. Harrison, M. J. Freedman et al., "Frenetic: a network programming language," in Proceedings of the 16th ACM SIGPLAN International Conference on Functional Programming (ICFP '11), pp. 279–291, ACM, Tokyo, Japan, September 2011.

## AUTHORS

**Amar Chaudhari Amar** is a graduate student from the University of Colorado Boulder with a major in Network Engineering. He received his Bachelor of Engineering degree in Information Technology, and currently works as a Site Reliability Engineer at LinkedIn.

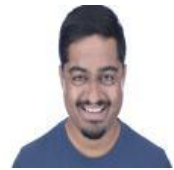

**Amrita Asthana Amrita** is a graduate student from the Interdisciplinary Telecom Program, University of Colorado Boulder. She received her bachelor's degree in Electronics and Communications Engineering, and currently works as a Network DevOps Engineer at Comcast.

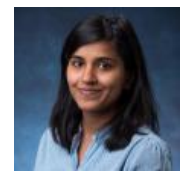

**Atharva Kaluskar Atharva** is a graduate student from the University of Colorado Boulder with a major in Network Engineering. He received his bachelor's degree in Electronics and Communications Engineering, and currently works as a Technical Solutions Engineer at Google.

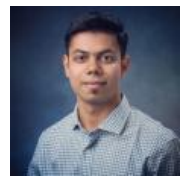

**Dewang Gedia Dewang** is a Ph.D. candidate at the Interdisciplinary Telecom Program, University of Colorado Boulder. He received his master's degree in Network Engineering, and has primary research focus in network functions virtualization and software- defined networks domain.

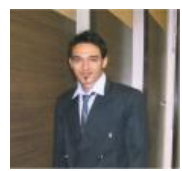

**Lakshay Karani Lakshay** is a graduate student from the Interdisciplinary Telecom Program, University of Colorado Boulder. He received his bachelor's degree in Electronics and Communications Engineering, and currently works as a Network Engineer at Charter Communications.

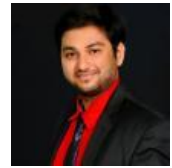

**Dr. Levi Perigo Dr. Perigo** is a Scholar in Residence and Professor of Network Engineering at the Interdisciplinary Telecom Program, University of Colorado Boulder. His interests are in a variety of internetworking technologies such as network automation, VoIP, IPv6, SDN/NFV, and next generation protocols. Currently, his research focuses on implementation and best practices for network automation, SDN, and NFV.

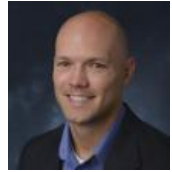

**Rahil Gandotra Rahil** is a Ph.D. student at the Interdisciplinary Telecom Program, University of Colorado Boulder. He received his bachelor's degree in Telecommunications Engineering, and has primary research interests in next-generation networking focusing on software-defined networking, network functions virtualization, and energy-efficient networking.

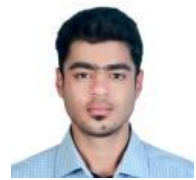

**Sapna Gangwar Sapna** is a graduate student from the Interdisciplinary Telecom Program, University of Colorado Boulder. She received her bachelor's degree in Electronics and Communications Engineering, and currently works as a Technical Solutions Engineer at Arista Networks.

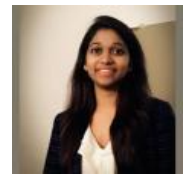